\documentclass[11pt,nofootinbib,floatfix]{revtex4}
\usepackage{mathrsfs}
\usepackage{graphicx}
\usepackage{amsmath}
\usepackage{amsfonts}
\usepackage{amssymb}
\usepackage{color}
\usepackage{hyperref}

\usepackage{epsfig}
\usepackage{CJK}
\usepackage{eepic}
\usepackage{bbm}
\usepackage{dcolumn}
\usepackage{bm}
\usepackage{ulem}
\usepackage{multirow}
\usepackage[caption=false]{subfig}

\usepackage{tikz}
\usetikzlibrary{graphs}
\usepackage{minipage-marginpar}

\newcommand{\omits}[1]{}

\def\bc{\begin{center}}
\def\nno{\nonumber}
\def\ec{\end{center}}
\def\be{\begin{eqnarray}}
\def\ee{\end{eqnarray}}

\definecolor{dyellow}{rgb}{1.,0.8,.0}
\definecolor{myblue}{rgb}{.1,.1,.7}
\definecolor{dcyan}{rgb}{.0,.6,.6}
\definecolor{cyan}{rgb}{0.4,1.0,1.0}
\definecolor{dmagenta}{rgb}{0.6,0.0,0.6}
\definecolor{brown}{rgb}{0.6,0.2,0.}
\definecolor{darkblue}{rgb}{.0,.0,0.5}
\definecolor{darkred}{rgb}{0.75,0.0,0.0}
\definecolor{orange}{rgb}{1.,.6,.0}
\definecolor{dorange}{rgb}{0.8,.4,.0}
\definecolor{green}{rgb}{0.0,1.0,0.0}
\definecolor{darkgreen}{rgb}{0.0,0.6,0.0}
\definecolor{purple}{rgb}{.4,.0,.4}
\definecolor{lightgrey}{rgb}{0.7, 0.7, 0.7}
\definecolor{grey}{rgb}{0.4, 0.4, 0.4}

\def\be{\begin{equation}}
\def\ee{\end{equation}}
\def\bea{\begin{eqnarray}}
\def\eea{\end{eqnarray}}

\def\>{\rangle} 
\def\<{\langle} 

\begin{document}


\title{Island of acoustic black hole in Schwarzschild spacetime}

\author{Yu-Ye Cheng$^{1}$} \email{chengyy53@mail2.sysu.edu.cn}
\author{Jia-Rui Sun$^{1}$} \email{sunjiarui@mail.sysu.edu.cn}

\affiliation{${}^1$School of Physics and Astronomy, Sun Yat-Sen University, Guangzhou 510275, China}



\begin{abstract}
We study an analogue information paradox in acoustic black holes which are emerged from the superfluid surrounding a Schwarzschild black hole. The resulting acoustic black hole contains both acoustic horizons and optical horizon, with analogue Hawking radiation, i.e. phonons, emitted from the outer acoustic horizon. By using the island formula, we calculate the entanglement entropy of analogue Hawking radiation of the acoustic black hole in both non-extremal and extremal cases. In the non-extremal case, the entanglement entropy of phonons follow the Page curve due to the emergence of islands, and it is approximately proportional to the area of the acoustic horizon at late time. While in the extremal case, the entanglement entropy of phonons diverges, leading to an ill-defined Page time. Our study verifies the unitarity of the analogue gravity system, and provides further insight into the connection between the entanglement entropy and the causal structure of spacetime.
\end{abstract}


\maketitle
\newpage
\tableofcontents

\section{Introduction}
The black hole information paradox, which argues that the evaporation of black holes may conflict with the unitarity in quantum mechanics~\cite{Hawking:1976ra,Hawking:1974rv,Hawking:1975vcx}, has been a central problem in gravitational physics for decades. The original calculations of Hawking suggested that black holes will emit thermal radiation with monotonically increasing entanglement entropy, which implied that information is lost as the black hole evaporates. However, the unitarity of quantum mechanics demands that the black hole and its radiation should remain in a pure state even at the end of evaporation. To avoid the information paradox in the semiclassical gravity, Page argued that during a unitary black hole evaporation process, the entanglement entropy of the radiation should increase first, then saturate, and eventually decrease, which satisfies the so-called Page curve~\cite{Page:1993wv}. Then the key point to solve this paradox is to correctly calculate the entanglement entropy of the radiation which obeys the Page curve.

Recently, important progresses have been made in solving the black hole information paradox at the semiclassical level, based on the holographic calculation of entanglement entropy, i.e., the Ryu-Takayanagi (RT) formula~\cite{Ryu:2006bv,Hubeny:2007xt}, and its quantum extension, i.e., the quantum extremal surface (QES) formula~\cite{Faulkner:2013ana}. The new approach proposed is called the island formula, which was shown can successfully reproduce the Page curve by calculating the entanglement entropy of Hawking radiation during black hole evaporation \cite{Penington:2019npb,Almheiri:2019psf,Almheiri:2019hni,Almheiri:2020cfm}. The formula was later confirmed by using the replica method for the gravitational path integral, where the emergence of new saddles, known as replica wormholes, lead to islands, which are regions located inside the black hole and belong to part of the entanglement wedge of the external radiation region~\cite{Penington:2019kki,Almheiri:2019qdq,Yu:2025tid}. The island formula explicitly states the transition in entanglement entropy, that leads to the Page curve. Explicitly, in the initial no-island phase, the entropy grows monotonically with time. After a critical point (the Page time), a new island phase dominates and makes the entropy stable. This transition ensures that the total entanglement entropy of the system follows the Page curve. The island formula can be explicitly expressed as
\be
  S_R = {\rm min}     \left\{  {\rm ext}  \left[     \frac{ {\rm Area} (\partial I) }{4 \tilde{G}_N}    +    S_{\rm matter}^{\rm finite}(I\cup R)    \right]\right\},
\ee
where $R$ is the radiation region and $\partial I$ is the boundary of the island, and the area-like divergences of the entanglement entropy of matter fields that depend on the short-distance cutoff have been absorbed into the renormalized Newton’s constant $\tilde{G}_N$~\cite{Bombelli:1986rw,Srednicki:1993im,Susskind:1994sm}. So far, the island formula has been successfully applied to resolve the information paradox in two dimensional gravity and has subsequently been generalized to other lower and higher dimensional black holes, see, e.g., \cite{Hollowood:2020cou,Goto:2020wnk,Anegawa:2020ezn,Gautason:2020tmk,Hartman:2020swn,
Wang:2021mqq,Wang1:2021woy,Li:2021lfo,Almheiri:2019psy,He:2021mst,Alishahiha:2020qza,
Hashimoto:2020cas,Krishnan:2020oun,Hu:2022zgy,Gan:2022jay,Du:2022vvg,Ling:2020laa,
Wang:2021woy,Miao:2023unv,Guo:2023fly,Guo:2023gfa,Kim:2021gzd,Tong:2023nvi,Du:2025kcx,Ahn:2021chg,
HosseiniMansoori:2022hok,RoyChowdhury:2022awr,Yu:2024fks,Espindola:2022fqb}. 

In the present paper, we will instead investigate the information paradox in analogue gravity. The motivations are the follows. Firstly, the typical feature of analogue gravity is the emergence of curved geometry, especially the horizon, which can produce analogue Hawking radiation composed of quantized phonons~\cite{Barcelo:2005fc}. Therefore, it was originally proposed to mimic curved geometries and quantum effects in curved spacetime. For example, Unruh first proposed the acoustic black hole from nonrelativistic fluids in flat spacetime to mimic the near horizon geometry and Hawking radiation of black hole~\cite{Unruh:1980cg}. Since then, analogue gravity have been successfully realized in various condensed matter systems such as the Bose-Einstein condensation~\cite{Lahav:2009wx}, the optical media~\cite{Drori:2018ivu} and many others (see the review~\cite{Barcelo:2005fc} on analogue gravity models). The advantage of analogue gravity is that they provide alternative experimental tools in lab for simulating quantum effects of in curved spacetime like the Hawking radiation of black holes~\cite{Steinhauer:2015saa}, which are inaccessible by current astrophysical observations due to technological limitations and precision requirements.\footnote{There are also relevant studies discussing the simulation of physical phenomena such as cosmic inflation and particle pair production through analogue gravity~\cite{Fischer:2004bf,Fedichev:2003id,Cha:2016esj}.} Secondly, the analogue gravity is not merely an analogy. Using the fluid/gravity duality, it was first sown in~\cite{Ge:2015uaa} that the acoustic black hole formed in fluid at the radial cutoff surface can be mapped to a real black brane in asymptotically AdS spacetime at the dynamical level, the Hawking-like temperature of the acoustic black hole is connected to the Hawking temperture of the AdS black brane, and moreover, the phonon is dual to the sound channel of the quasinormal mode in the bulk AdS black brane. The dynamical connection between acoustic black hole in fluid and black hole has also been extended to the black D3 brane in flat spacetime~\cite{Yu:2017bnu}. Thirdly, apart from systems in flat spacetime, analogue gravity can also appear in curved spacetime, e.g., acoustic black holes have been extended in general curved spacetimes from relativistic Gross-Pitaevskii (GP) theory and Yang-Mills (YM) theory, where a real black hole is surrounded by an acoustic horizon, which is equivalent to a `dumb hole' from which neither light nor sound can escape~\cite{Ge:2019our}. Thus the analogue gravity in curved spacetime can be used to describe real black holes surrounded by environments such as some fluid accreting on to a black hole~\cite{Das:2004zm}. Later works have discussed effects such as black hole shadows, quasinormal modes, and analogue Hawking radiation in this class of analogue gravity \cite{Wang:2019zqw,Guo:2020blq,Ling:2021vgk,Qiao:2021trw,Wang:2022gbl,Parvizi:2023foz}. 

Based on these specific properties of analogue gravity, it is natural to ask whether the analogue Hawking radiation in acoustic black holes can preserve unitarity as well? To answer this question, we will consider acoustic black holes formed in a four-dimensional Schwarzschild black hole background. Due to the presence of multiple acoustic horizons, the acoustic black hole can be classified into non-extremal and extremal cases, similar to the Reissner-Nordström (RN) black hole, although the singularity remains spacelike in this case. We utilize the island formula to compute the entanglement entropy of analogue Hawking radiation from acoustic black holes in different cases and reconstruct the corresponding Page curves. In non-extremal case, we find that in the absence of the island contribution, the entanglement entropy of radiative phonons exhibits a linear time dependence in the late-time regime. Subsequently, we calculate the entanglement entropy when the radiation region is close to and far from the acoustic horizon, and reproduce the Page curve. In the extremal case, our analysis shows that, although a symmetric Penrose diagram exists, the vanishing surface gravity still leads to a divergence of entanglement entropy. This result is different from previous works on extremal black holes \cite{Kim:2021gzd,Ahn:2021chg,HosseiniMansoori:2022hok,Tong:2023nvi}, which indicates that the divergence of entanglement entropy is caused by the Cauchy horizon terminating at a timelike singularity. Although this makes the Page time ill-defined, the total entropy will be finite when an island exists. This suggests that information may be closely related to the causal structure of spacetime.

The paper is organized as follows. In sec.~\ref{review} we give a brief review on basic properties of acoustic black holes in curved spacetime. Then in sec.~\ref{non-extremal} and sec.~\ref{extremal} we calculate the entanglement entropy of analogue Hawking radiation using the island formula both in the nonextremal and extremal Schwarzschild acoustic black hole, respectively. The Page curve will be investigated in sec.~\ref{pagecurve}. Finally, we give conclusions and discussions in sec.~\ref{con}.
\section{Review of acoustic black hole in curved spacetime} \label{review}
In this section, we give a brief review of the acoustic black hole in Schwarzschild black hole spacetime~\cite{Ge:2019our}. The acoustic black hole in a general curved spacetime can be constructed from the relativistic GP theory, the action for a complex scalar field $\varphi$ in this theory is
\be
 S  =\int d^{4} x \sqrt{-g} \left(   \left| \partial_\mu \varphi \right |^2   +    m^2\left|\varphi\right|^2    -    \frac{b}{2} \left|\varphi\right|^4   \right),
\ee
where $b$ is a constant and $m^2$ is a parameter depending on temperature as $m^2 \sim \left(T-T_c\right)$. The EoM for the scalar field $\varphi$ is
\be\label{eq:KG}
      \square \varphi \,+\, m^2\varphi \, - b\left|\varphi\right|^2\varphi = 0.
\ee
One can choose a fixed static background spacetime $ds_{\rm bg}^2$ as
\be
  ds_{\rm bg}^2 = g_{tt}dt^2 + g_{rr}dr^2 + g_{\theta\theta}d\theta^2 + g_{\phi\phi}d\phi^2.
\ee
By using the Madelung representation $\varphi = \sqrt{\rho(\vec{x},t)} e^{i \theta (\vec{x},t)}$, and considering perturbations via $\rho=\rho_0+\rho_1$ and $\theta=\theta_0+\theta_1$, Eq.~\eqref{eq:KG} reduces to those for the background and phase fluctuations, respectively
\bea
  b \rho_0 = m^2 - g_{\mu\nu} \partial_\mu \theta_0 \partial_\nu \theta_0 &=& m^2 -v_\mu v^\mu,\label{bgd eq} \\
  \frac{1}{\sqrt{-\cal G}}\partial_\mu \left(\sqrt{-\cal G}{\cal G}^{\mu\nu }\partial_\mu\theta_1\right) &=& 0.
\eea
where $v_\mu = \{-\partial_0\theta_0,\partial_i\theta_0\}$, and the effective metric ${\cal G}_{\mu\nu}$ for the phase fluctuations can be obtained as
\be
      {\cal G}_{\mu\nu} = \frac{c_s}{\sqrt{c_s^2-v_\mu v^\mu}} \begin{pmatrix}
        g_{tt}(c_s^2-v_iv^i) & \vdots &\qquad -v_i v_t \qquad\\
        \cdots\cdots\cdots\cdots& \cdot & \cdots\cdots\cdots\cdots\cdots\cdots \cdots\cdots  \\
        -v_iv_t& \vdots & g_{ij} \left(c_s^2-v_\mu v^\mu\right)+v_iv_j\\
    \end{pmatrix}.
\ee
The speed of sound is defined as  $c_s^2 \equiv \frac{b\rho_0}{2}$. The effective metric ${\cal G}_{\mu\nu}$ depends on the background spacetime and the four-velocity of the fluid. By considering $v_r\neq0,v_t\neq0,v_a=0$ for $(a=\theta,\phi)$, and $g_{tt}g_{rr}=-1$, and taking the coordinate transformation $dt \to  dt - \frac{v_tv_r}{g_{tt}\left(c_s^2-v_rv^r\right)}dr$, the effective metric can be rewritten as
\be
 ds^2  = c_s \left(\frac{ c_s^2 -v_rv^r }{ \sqrt{c_s^2 -v_\mu v^\mu} }g_{tt} dt^2 + \frac{c_s^2 \sqrt{c_s^2 -v_\mu v^\mu }}{c_s^2 -v_rv^r} g_{rr}dr^2 + \sqrt{c_s^2 -v_\mu v^\mu}g_{\theta\theta}d\theta^2 + \sqrt{c_s^2 -v_\mu v^\mu}g_{\phi\phi}d\phi^2  \right) .
\ee

Now let us choose the background spacetime to be a Schwarzschild black hole
\be
    ds_{\rm bg}^2 = -f(r)dt^2 + \frac{1}{f(r)}dr^2 + r^2(d\theta^2+\sin^2\theta d\phi^2),
\ee
where $f(r) = 1 - \frac{r_h}{r}$ and $r_h$ is the horizon radius. Considering a vortex that is falling radially from infinity towards the black hole. The radial velocity $v_r$ should not be smaller than the escape velocity at Schwarzschild coordinate radius $r$, and can be set as $v_r\sim\sqrt{2M\xi/r}$. Here $\xi$ is a tuning parameter. After rescaling $m^2\to \frac{m^2}{2c_s^2}$ as well as $v_\mu v^\mu \to \frac{v_\mu v^\mu}{2c_s^2}$, Eq.~\eqref{bgd eq} becomes $v_\mu v^\mu = m^2-1$. At the critical temperature $T_c$, the  mass parameter $m^2$ vanishes, leading to the condition $v_\mu v^\mu=-1$. Under these conditions, the effective metric can be rewritten as
\bea
  &ds^2& = \sqrt{3}c_s^2 \left(-{\cal F}(r )dt^2 + \frac{1}{{\cal F}(r)}dr^2+r^2 \left(d\theta^2+\sin^2\theta d\phi^2\right)\right) , \nno\\
  &{\rm with}& \; {\cal F}(r) = \left(1-\frac{r_h}{r} \right) \bigg( 1-\xi\frac{r_h}{r}\left(1-\frac{r_h}{r}\right) \bigg).
\eea
For convenience, we set $c_s^2=1/\sqrt{3}$.

The metric function ${\cal F}(r)=0$ has three roots: the optical event horizon $r_h$ and two acoustic horizons $r_{\pm}$, in which the acoustic horizons are given by
\be
  r_{\pm} = \frac{r_h}{2}\left(\xi \pm \sqrt{\xi^2-4\xi }\right),
\ee
where the condition $\xi \geq 4$ is required to ensure the existence of two acoustic horizons. These horizons define four distinct regions within the spacetime:
\begin{itemize}
    \item The region inside the optical horizon $r<r_h$,
    \item Two regions $r_h < r < r_-$ and $r_- < r <r_+$  where sound cannot escape,
    \item The region outside the outer acoustic horizon $r>r_+$ where both light and sound can escape.
\end{itemize}
For $\xi > 4$, the horizons are distinct, satisfying the relation $r_h < r_- < r_+$. When $\xi =4 $, the two acoustic horizons coincide, resulting in a single horizon at $r_+=r_-=2r_h$. The acoustic black hole becomes extremal.

The Hawking temperature and the area entropy at the outer horizon are
\be
  T_H = \frac{1}{\beta} =   \frac{{\cal F}'(r_+)}{4\pi} = \frac{(r_+-r_h)(r_+-r_-)}{4\pi r_+^3} = \frac{\kappa_+}{2\pi }, \quad S = \frac{A}{4G_N} = \frac{\pi r_+^2}{G_N},
\ee
where $\kappa_+$ is the surface gravity of the outer horizon $r_+$. The surface gravity of different horizons could be derived from the tortoise coordinate
\bea
  r^* &=& r     + \frac{r_-^3}{(r_--r_+)(r_--r_h)}\log( \frac{r}{r_-}-1)    +\frac{r_+^3}{(r_+-r_-)(r_+-r_h)}\log( \frac{r}{r_+}-1)  \nonumber\\
    & \quad&      \;\;\,+\;\frac{r_h^3}{(r_h-r_-)(r_h-r_+)}\log( \frac{r}{r_h}-1) \nonumber\\
    &=&                 r+\frac{1}{2\kappa_-}\log (\frac{r}{r_-}-1)+ \frac{1}{2\kappa_+}\log(\frac{r}{r_+}-1)+{\frac{1}{2\kappa_h}}\log(\frac{r}{r_h}-1).
\eea
We can further define the Kruskal coordinates as
\bea\label{eq:Kruskal}
    ds^2 &=& -\Omega^2 dU dV + d\Omega^2, \\
    U  &=& -e^{-\kappa_+(t-r^*)}, \quad V = e^{\kappa_+(t+r^*)}.
\eea
with the conformal factor
\be
  \Omega^2 = \frac{{\cal F}(r)}{\kappa_+^2}e^{-2\kappa_+r^*}.
\ee

In the extremal case $\xi = 4$ , the acoustic horizon becomes $2r_h$ and the metric function ${\cal F}(r)$ becomes $\frac{(r-r_h)(r-r_+)^2}{r^3}$. The tortoise coordinate now is
\be\label{ex_rs}
  r^*  =r-\frac{r_+^3}{(r-r_+)(r_+-r_h)}+\frac{r_h^3}{(r_h-r_+)^2}\log\left(\frac{r-r_h}{r_h}\right)+\frac{r_+^2( 2r_+ -3r_h )}{(r_h-r_+)^2}\log\left(\frac{r-r_+}{r_+}\right) .
\ee
The surface gravity $\kappa_+=\kappa_-$ now vanishes, and the Hawking temperature $T_H=0$. Therefore, we approach the result of calculation by taking the limit $\kappa_+\to 0$, as in previous studies \cite{Kim:2021gzd,Ahn:2021chg,HosseiniMansoori:2022hok,Tong:2023nvi}.
\section{The entanglement entropy in non-extremal case}\label{non-extremal}
In this section, we will calculate the entanglement entropy of the analogue Hawking radiation for the non-extremal Schwarzschild acoustic black hole. We will first discuss the entanglement entropy in the absence of islands, then analyze two cases that the radiation region is close to and far from the outer horizon when islands exist.
\subsection{Without islands}
\begin{figure}[htbp]
    \includegraphics[width=0.8\textwidth,clip]{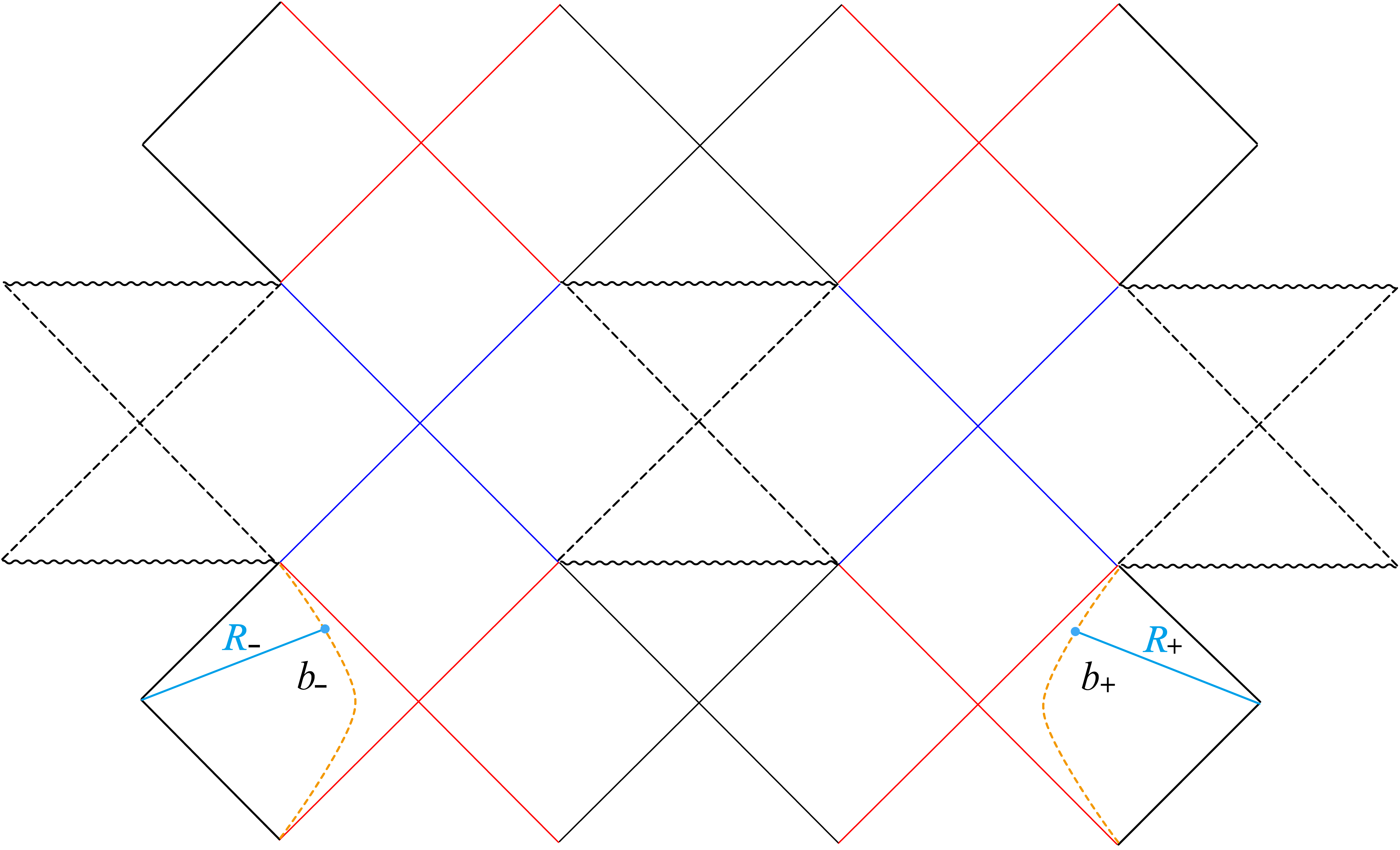}
    \caption{The part of Penrose diagram for a non-extremal Schwarzschild acoustic black hole without islands. The outer and inner acoustic horizons $r_+,r_-$ are denoted by red lines and blue lines. The optical horizon $r_h$ is the dashed line. The black line represents the null infinity. In the absence of islands, the total entanglement entropy is determined by boundaries $b_\pm$ of the radiation region $R=R_+\cup R_-$. The Penrose diagram can be infinitely continued in both the vertical and horizontal directions.}
    \label{fig:penrose1}
\end{figure}

In the absence of islands, there are two points of the entanglement regions located in the right wedge $R_+$ and the left wedge $R_-$. We set the coordinates of these points as $b_+=(t_b,b)$ and $b_-=(-t_b+i\beta/2,b)$, see Figure \ref{fig:penrose1}. It can be found that the Penrose diagram of an acoustic Schwarzschild black hole is similar to that of a RN black hole rotated by 90 degrees, with the addition of a region for the outer acoustic horizon. When the distance between the two points is large, we can calculate the entanglement entropy of the analogue Hawking radiation composed of quantized phonons by assuming the s-wave approximation \cite{Hashimoto:2020cas}
\be\label{noisland1}
S_{\rm phonon} = -I(R_+,R_-) = \frac{c}{3} \log d(b_+,b_-),
\ee
where $c$ is the central charge and the mutual information is be given by the distance between the boundaries $d(b_+,b_-)$. Through a conformal transformation, we obtain the total entropy in Kruskal coordinates
\bea
S_{\rm gen}=S_{\rm phonon} &=& \frac{c}{6} \log \left[\Omega(b_+)\Omega(b_-)\left( {U(b_-)-U(b_+)} \right)\left( {V(b_+)-V(b_-)} \right)\right] \nonumber\\
               &=&\frac{c}{6} \log \left[ 4 \frac{{\cal F}(b)}{\kappa_+^2}\cosh^2(\kappa_+t_b ) \right].
\eea
At late time $t_b \gg b \,> r_+$, the result is approximated as
\bea
 S_{\rm gen}&\simeq& \frac{c}{3}\kappa_+ t_b \nonumber\\
    & =&  \frac{c}{12} \frac{\big(4-\xi+\sqrt{(\xi-4)\xi}\big)}{r_h\xi} t_b.
\eea
We find that the non-negativity of entropy also requires $\xi \geq 4$. In this non-island phase, the entanglement entropy of radiation diverges with time, which is similar to Hawking's original calculations. In the following section, we will consider the entanglement entropy with the island, which should lead to the Page curve after the Page time.

\subsection{The radiation region near the horizon}
\begin{figure}[ht]
    \includegraphics[width=0.8\textwidth]{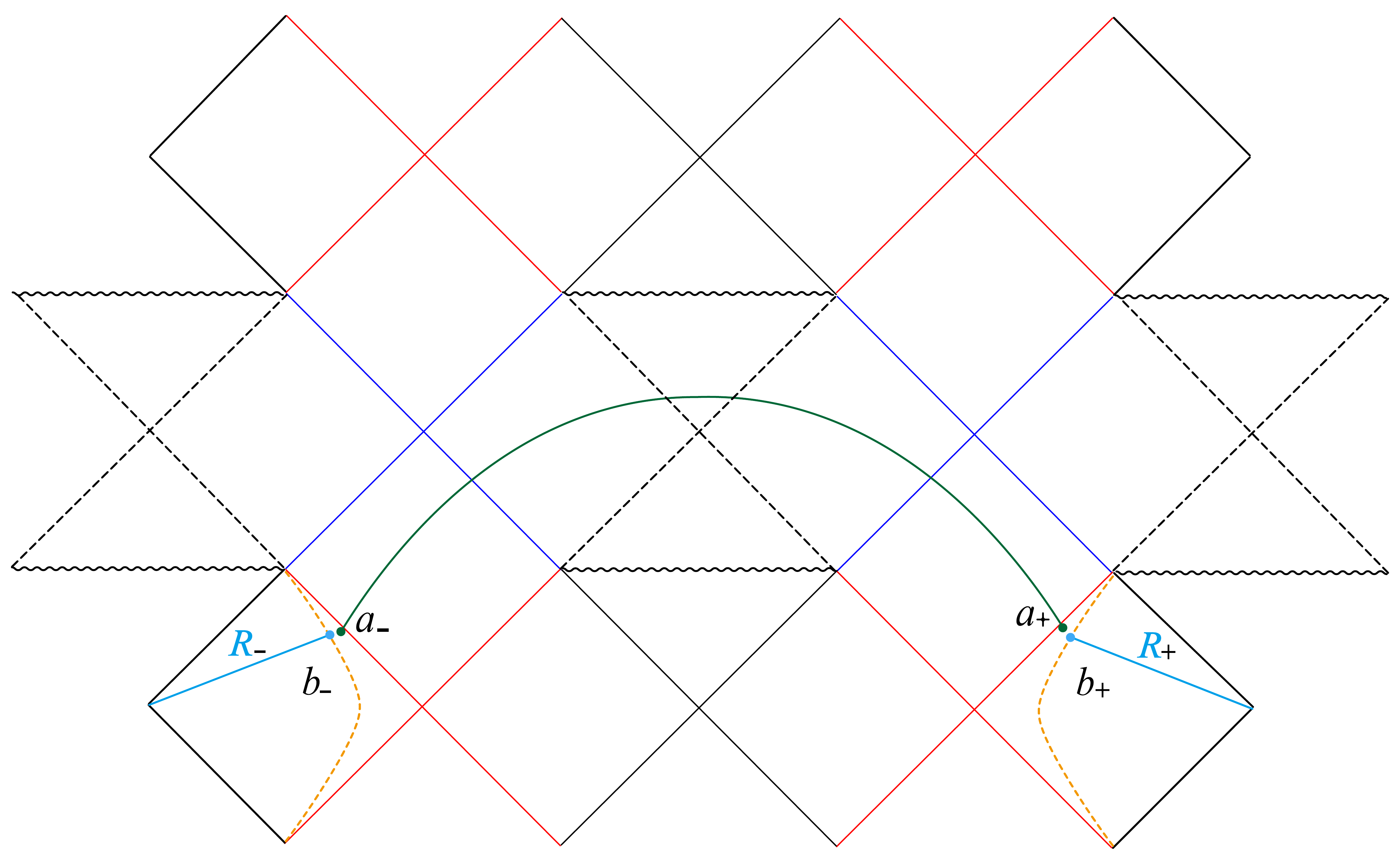}
    \caption{The part of Penrose diagram for a non-extremal  Schwarzschild acoustic black hole with island. The island extends between the right wedge and the left wedge (green line). The boundaries of the radiation region $R$ are denoted by $b_\pm$, and the boundaries of the island $I$ are denoted by $a_\pm$.}
    \label{fig:penrose 2}
\end{figure}

Now we will calculate the entanglement entropy with islands. The boundary of the radiation region is assumed to be very close to the acoustic horizon, i.e., $b-r_+\ll r_+$. For the case of a single island, its boundaries are set as $a_+=(t_a,a)$ and $a_-=(-t_a+i\beta/2,a)$, see Figure \ref{fig:penrose 2}. Since the distance between the island boundary and the radiation region boundary is sufficiently small, the finite part of entanglement entropy of the massless matter field can be evaluated as \cite{Casini:2005zv,Casini:2009sr}
\be\label{withisland near}
   S_{\rm phonon} = -2I(R_+,I) = -2\kappa c \frac{{\rm Area}}{L^2}.
\ee
Here $\kappa$ is a dimensionless constant that should not be confused with the surface gravity. The factor 2 comes from the fact that the entanglement entropy is calculated for both the left and right wedges. In the acoustic black hole, the geodesic distance between the boundary of region $I$ and that of region $R$ is analytically given by
\bea
   L &=& \frac{1}{\sqrt{r(r-c)(r-r_+)}}\bigg\{\frac{r(r-r_-)(r-r_+)}{\sqrt{r-r_h}}    -   \sqrt{\frac{(r-r_-)(r-r_+)r}{(r_h-r_-)r_+}}   \nonumber \\
     &\quad& \times\; \Big[(r_h-r_-)r_+\,E\left(\arcsin\left.A_1\,\right|\,A_2\right)- \;  (r_h^2-r_-r_++(r_-+r_+)r_h)\,F\left(\arcsin\left.A_1\,\right|\,A_2\right) \;   \nonumber \\
     &\quad& +\;(r_h-r_+)(r_h+r_-+r_+)\,\Pi( {A_2 r_+ }/r_h \,\,;\arcsin\left.A_1\,\right|\,A_2)\Big]\bigg\} \Big|_{a}^{b},
\eea
where $F(\phi|m),E(\phi|m),\Pi(n;\phi|m)$ are the elliptic integrals of the first kind, the second kind and the third kind, and $A_1=\sqrt{\frac{(r_--r_h)(r-r_+)}{(r-r_h)(r_--r_+) }},\; A_2= \frac{(r_+-r_-)r_h}{(r_h-r_-)r_+}$. Taking the limit $a-r_{+}\ll r_{+}$ and $b-r_{+}\ll r_{+}$, we obtain the short distance limit of above integral as
\be
  L \simeq  2 \sqrt{\frac{r_+^3}{(r_+-r_h)(r_+-r_-)}}\left(\sqrt{b-r_+}-\sqrt{a-r_+}\right) = 2\sqrt{\frac{1}{2\kappa_+}}\left(\sqrt{b-r_+}-\sqrt{a-r_+}\right).
\ee

Thus the total entanglement entropy of the radiation, i.e., the generalized entanglement entropy, is
\bea\label{eq:generalized entropy}
    S_{\rm gen} &=&  \frac{2\pi a^2}{G_N} - 8\kappa c \frac{\pi b^2}{L^2} \nonumber\\
                &=&  \frac{2\pi a^2}{G_N}-\frac{4\pi b^2\kappa c \kappa_+}{\left(\sqrt{b-r_{+}}-\sqrt{a-r_{+}}\right)^2} \nonumber \\
            &\simeq& \frac{2\pi a^2}{G_N}-\frac{4\pi r_+ \kappa c  \kappa_+}{\left(y-x\right)^2}
\eea
with new parameters $x=\sqrt{a-r_+}>0$ and $y=\sqrt{b-r_+}>0$. We take $x<y\ll1$ as the new near-horizon condition. According to the island formula, the actual entanglement entropy requires taking the minimum value among all extremal cases. Since the result is independent of time, we only need to take the derivative with respect to the position of the island, which is equivalent to
\be
  \frac{x}{y}(1-\frac{x}{y})^3 =\frac{  G_N\kappa c  \kappa_+}{r_+y^4}.
\ee
The above equation has an extremum point, provided that the right-hand side is smaller than the maximum value $27/256$ of the function \cite{Du:2022vvg,Tong:2023nvi}. Moreover, since the right-hand side is sufficiently small, we can approximately solve for the position of the island as
\be
  a = r_+ + \frac{(r_+G_N \kappa c \kappa_+)^2}{(b-r_+)^3},
\ee
which shows that there is a modification of order $O(G_N^2)$ compared to the position of the acoustic horizon, indicating that the island is located just slightly outside the horizon.
Substituting this into Eq.~\eqref{eq:generalized entropy},
\be\label{eq:close ee}
  S_{\rm gen}  \simeq  \frac{2\pi r_+^2}{G_N} - 4\pi \kappa c\frac{ r_+^2\kappa_+}{b-r_+},
\ee
in which the first term is twice the area entropy at acoustic horizon $r_+$, while the second term represents the effect of quantum corrections, and the resulting entanglement entropy is now finite and no longer diverges with time.

\subsection{The radiation region far from the horizon}
Next, we will consider the case where the radiation region is far from the acoustic horizon. Since the s-wave approximation remains valid, the entanglement entropy in this case is
\be\label{eq:ee far formula}
  S_{\rm phonon} = \frac{c}{3}\log \frac{d(a_+,a_-)d(b_+,b_-)d(a_+,b_+)d(a_-,b_-)}{d(a_+,b_-)d(a_-,b_+)}.
\ee
Working in the Kruskal coordinates, the generalized entanglement entropy is expressed as
\bea
  S_{\rm gen} & = &       \frac{2\pi a ^2}{G_N} + \frac{c}{6}\log \left[\frac{2^4{\cal F}(a){\cal F}(b)}{\kappa_+^4}\cosh^2(\kappa_+t_a)\cosh^2(\kappa_+t_b)\right] \nonumber \\
    & \quad &   +\;\frac{c}{3}\log \left\{ \frac{\cosh\left[\kappa_+\left({r_*(a)-r_*(b)}\right)\right]-\cosh\left[\kappa_+(t_a-t_b)\right]}{\cosh\left[\kappa_+\left({r_*(a)-r_*(b)}\right)\right]+\cosh\left[\kappa_+(t_a+t_b)\right]}\right\} .
\eea
We can further use the following equation for simplification:
\bea
    \cosh\left[\kappa_+\left({r_*(a)-r_*(b)}\right)\right]  = \frac{1}{2}
    \left[
        \sqrt{\frac{K(a)}{K(b)}}e^{\left[\kappa_+(a-b)\right]} + \sqrt{\frac{K(b)}{K(a)}}e^{\left[\kappa_+(b-a)\right]}
    \right],
\eea
where $K(r)=(r-r_+)(r-r_-)^{\frac{\kappa_+}{\kappa_-}}(r-r_h)^{\frac{\kappa_+}{\kappa_h}}$. For $a\sim r_+$, the first term can be ignored since the second term dominates.

Then we can take the late time limit~\cite{Tong:2023nvi}
\be
  \frac{1}{2}\sqrt{\frac{K(b)}{K(a)}}e^{\left[\kappa_+(b-a)\right]} \ll \cosh \left[\kappa_+\left(t_a+t_b \right)\right],
\ee
and the approximation
\be
  \cosh \left[\kappa_+\left(t_a-t_b \right)\right]\ll  \frac{1}{2}\sqrt{\frac{K(b)}{K(a)}} e^{\left[\kappa_+(b-a)\right]}.
\ee
Under these conditions, the entanglement entropy can be simplified as:
\bea
    S_{\rm gen} &=& \frac{2\pi a^2}{G_N} + \frac{c}{6}\log \left[\frac{2^4{\cal F}(a){\cal F}(b)}{\kappa_+^4}\cosh^2(\kappa_+t_a)\cosh^2(\kappa_+t_b)\right]  \nonumber\\
    &\quad& -\;\frac{c}{3}\log\left[2\sqrt{\frac{K(a)}{K(b)}}e^{\kappa_+(a-b)}\cosh[\kappa_+{(t_a+t_b)}]\right] - \frac{2c}{3}\sqrt{\frac{K(a)}{K(b)}}e^{\kappa_+(a-b)}\cosh[\kappa_+{(t_a-t_b)}] \nonumber\\
    &\approx&  \frac{2\pi a^2}{G_N} + \frac{c}{6}\log\left[\frac{{\cal F}(a){\cal F}(b)K(b)}{K(a)} e^{2\kappa_+(b-a)}\right] - \frac{2c}{3}\sqrt{\frac{K(a)}{K(b)}}e^{\kappa_+(a-b)}\cosh[\kappa_+{(t_a-t_b)}].
\eea

To obtain the extremum, we first take the derivative with respect to the location $a$, which gives
\be
  a \simeq r_+    +   \frac{\left(cG_N\right)^2\left(\frac{r_+-r_-}{b-r_h}\right)^{\frac{\kappa_+}{\kappa_h}}\left(\frac{r_+-r_-}{b-r_-}\right)^{\frac{\kappa_+}{\kappa_-}}}{144\pi^2(b-r_+)r_+^2}e^{2\kappa_+(r_+-b)}\cosh^2[\kappa_+(t_a-t_b)].
\ee
It is clear that the position of the island is also slightly outside the horizon, and the entanglement entropy can be approximated as
\bea
     S_{\rm gen} &\approx& \frac{2\pi r_+^2}{G_N} + \frac{c}{6}\log\left[\frac{4{\cal F}(b)K(b)}{\kappa_+^4r_+^3}  (r_+-r_h)^{1-\frac{\kappa_+}{\kappa_h}}  (r_+-r_-)^{1-\frac{\kappa_+}{\kappa_-}}  e^{\kappa_+(b-r_+)}\right] \nonumber \\
                  &\quad& -\; \frac{c^2G_N\left(\frac{r_+-r_-}{b-r_h}\right)^{\frac{\kappa_+}{\kappa_h}}\left(\frac{r_+-r_-}{b-r_-}\right)^{\frac{\kappa_+}{\kappa_-}}}{36\pi r_+ (r_+-b)}\left(1+2e^{\kappa_+(r_+-b)}\right)\cosh^2[\kappa_+{(t_a-t_b)}].
\eea
Besides, $\frac{\partial S_{\rm gen}}{\partial t}$ gives $t_a=t_b$. Then, ignoring the higher order terms, the entanglement entropy of radiation with the island is
\bea\label{eq:far ee}
   S_{\rm gen} = \frac{2\pi r_+^2}{G_N} +  \frac{c}{6}\log\left[\frac{4{\cal F}(b)K(b)}{\kappa_+^4r_+^3}  (r_+-r_h)^{1-\frac{\kappa_+}{\kappa_h}}  (r_+-r_-)^{1-\frac{\kappa_+}{\kappa_-}}  e^{\kappa_+(b-r_+)}\right],
\eea
which shows that when the radiation region is far from the acoustic horizon, the entanglement entropy of radiation eventually reaches a stable value. This outcome is consistent with the case where the radiation region is close to the acoustic horizon, as both results include an area term and a quantum correction term. In both cases, the entanglement entropy initially grows with time before transitioning to this stable value at the Page time. However, as we will show in the next section, the situation is different for extremal black holes. Although we can still compute the entanglement entropy with the existence of an island, the Page time cannot be well defined.

\section{The entanglement entropy in the extremal case}\label{extremal}
In this section, we will focus on the extremal acoustic black hole in Schwarzschild spacetime, where the two acoustic horizons coincide at $r_+=r_-=2r_h$. Previous studies \cite{Kim:2021gzd,Ahn:2021chg,Tong:2023nvi} have shown that the extremal case is significantly different from the non-extremal case, particularly regarding the Penrose diagram. This is because the extremal geometry is not obtained by taking the continuous limit of the non-extremal case. We will adopt the same approach to calculate the entanglement entropy, both without and with islands. Furthermore, since the entanglement entropy depends on the specific position of the radiation region, we will begin with the case where the region is close to the horizon.
\subsection{Without islands}
\begin{figure}[htbp]
    \centering 
    \begin{minipage}[b]{0.48\columnwidth}
        \centering
        \includegraphics[width=\linewidth]{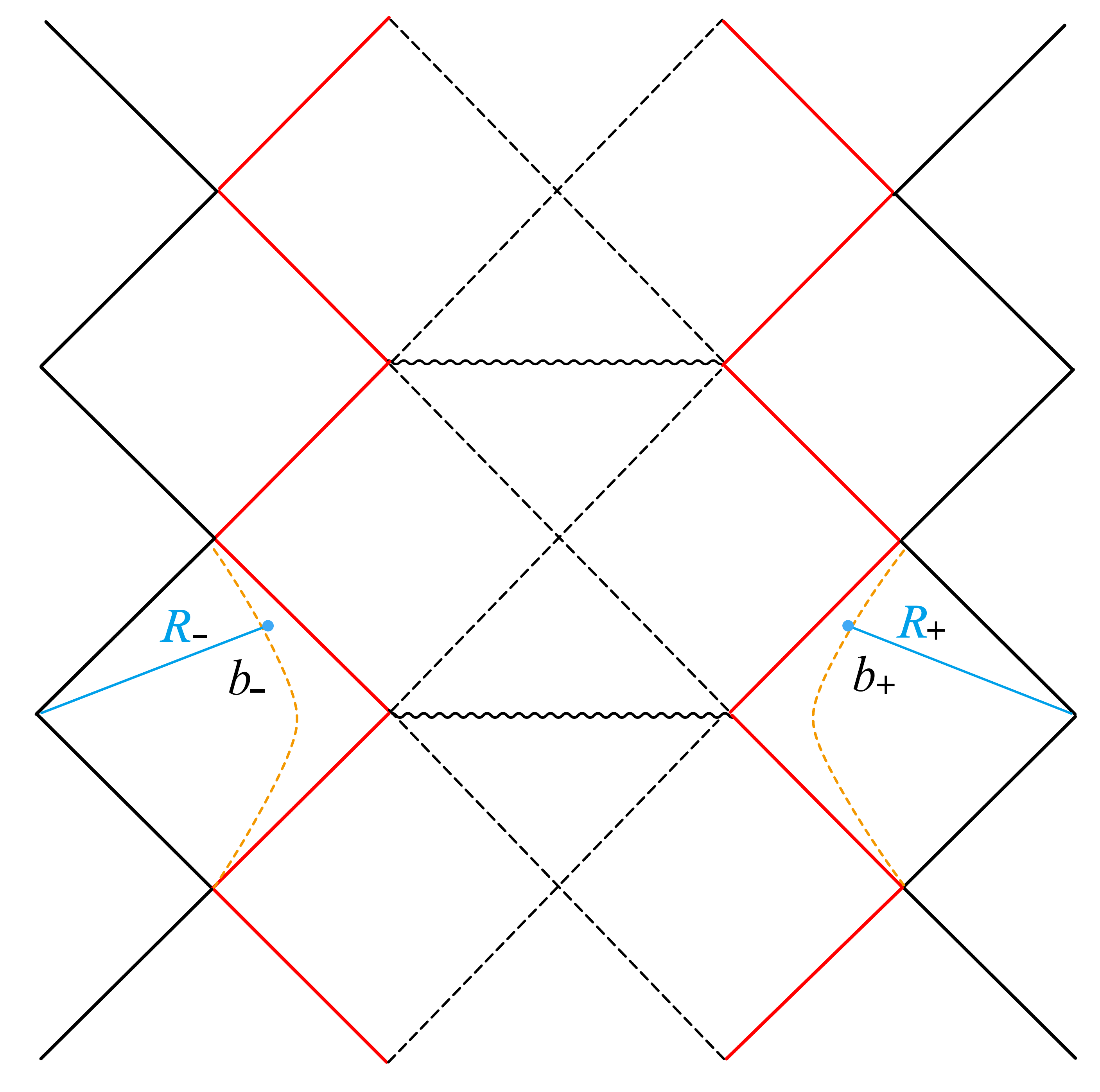}
        \par\vspace{3pt}
        (a) Without island
    \end{minipage}
    \hfill 
    \begin{minipage}[b]{0.48\columnwidth}
        \centering
        \includegraphics[width=\linewidth]{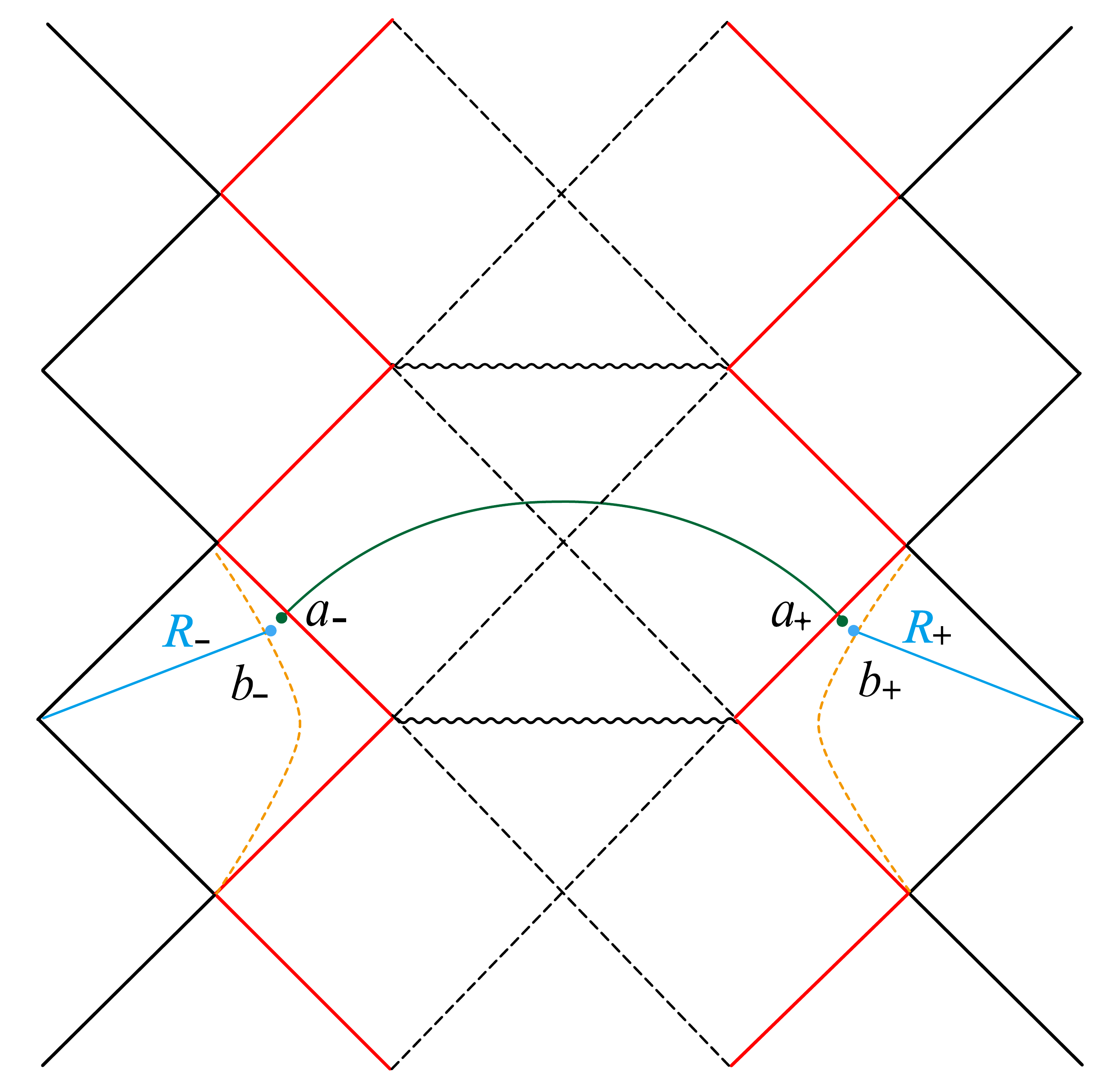}
        \par\vspace{3pt}
        (b) With island
    \end{minipage}
    
    \caption{The Penrose diagram of the Schwarzschild acoustic black hole in the extremal case with and without island.}
    \label{fig:penrose3/4}
\end{figure}

In contrast to previous studies \cite{Kim:2021gzd,Ahn:2021chg,Tong:2023nvi} where the Penrose diagram of the spacetime reduces to a one-sided structure due to a timelike singularity blocking the propagation of radiation, our case features a symmetric Penrose diagram because of the acoustic horizon. Consequently, as showed in Figure \ref{fig:penrose3/4}, there are two boundaries of the radiation region, denoted as $b_+=(t_b,b)$ and $b_-=(-t_b+i\beta/2,b)$, located at the right and left wedges. By using Eq.~\eqref{noisland1}, the entanglement entropy of Hawking radiation is given by
\bea
  S_{\rm gen}=S_{\rm phonon} &=& \lim_{\kappa_+\rightarrow0} \frac{c}{3}\log d(b_+,b_-) \nonumber \\
    &=&   \lim_{\kappa_+\rightarrow0} \frac{c}{3} \log \left(\frac{2}{\kappa_+}\cosh(\kappa_+t)\sqrt{{\cal F}(b)}\right),
\eea
in which $r^*$ is defined in Eq.~\eqref{ex_rs} and ${\cal F}(b)={(b-r_h)(b-r_+)^2}/{b^3}$. It can be seen that the entanglement entropy diverges as $\kappa_+\to 0$. While this divergence in the absence of islands implies that the Page time cannot be well-defined, the presence of islands still allows us to derive an exact result for the entanglement entropy.

\subsection{The radiation region near the horizon}
We set the boundaries of islands as $a_+=(t_a,a)$ and $a_-=(-t_a+i\beta/2,a)$,
and consider the case where the radiation region is near the acoustic horizon first. Assuming the boundaries of the island are outside and near the horizon such that $a-r_+\ll b-r_+ \ll r_+$, the geodesic distance is
\be
  L= \int_{a}^{b} \sqrt{\frac{r^3}{(r-r_h)(r-r_+)^2}}dr \simeq \frac{2r_+^{\frac{3}{2}}\log\left(\frac{x}{y}\right)}{\sqrt{r_+-r_h}}.
\ee
By using Eq.~\eqref{withisland near}, the generalized entropy is
\bea \label{Sgenex1}
  S_{\rm gen} &=&  \frac{2\pi a^2}{G_N} - c\kappa\frac{8\pi b^2}{L^2} \nonumber\\
      &=& \frac{2\pi a^2}{G_N} -  c\kappa\frac{2\pi b^2(r_+-r_h)}{r_+^3 \log\left(\frac{x}{y}\right)^2},
\eea
where the parameters $x=\sqrt{a-r_+}>0$ and $y=\sqrt{b-r_+}>0$, and $x\ll y\ll 1$. Before finding the extremal solutions of Eq.~(\ref{Sgenex1}), we first determine whether an island solution exists by analyzing the relevant conditions. From the extremum condition $\frac{\partial S_{\rm gen}}{\partial a}$, one obtains
\be
  \left(\frac{y^2}{x^2}\log\left(\frac{x^2}{y^2}\right)^3\right) =  \frac{2 c \kappa b^2 G_N (r_+-r_h)}{r_+^5 x^2},
\ee
further defining $z\equiv \frac{y^2}{x^2}$, the above equation becomes
\be
  z \left(\log \frac{1}{z}\right)^3 =  \frac{2 c \kappa b^2 G_N (r_+-r_h)}{r_+^5 x^2} \equiv F_0.
\ee
To ensure the existence of the island solution, the constant $F_0$ must be less than the maximum value of the function $F=z(\log\frac{1}{z})^3$ in the interval $z\in(0,1)$. In this case, the line of $F_0$ intersects the graph of the function $F$ at two points, which correspond to two solutions $z_1$ and $z_2$. And $z_1$ corresponds to the island solution, which implies the constraint
\be
  \frac{27}{e^3} > \frac{2 c \kappa b^2 G_N (r_+-r_h)}{r_+^5 x^2},
\ee
which is the special case of \cite{Tong:2023nvi} for $d=3$. When the constraint is violated, there would be no nontrivial island solution.

\begin{figure}[ht]
    \includegraphics[width=0.65\textwidth]{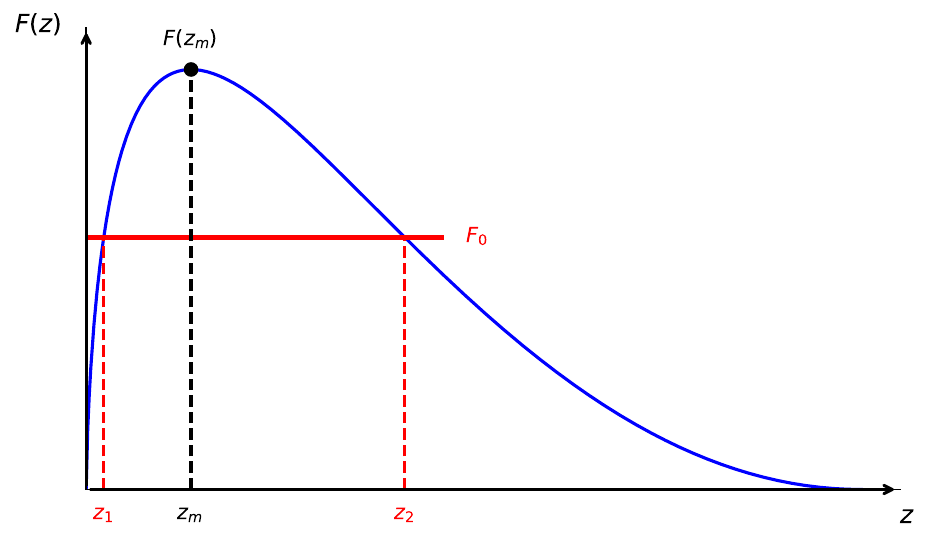}
    \label{Figure:xy1}
    \caption{The diagram of function $F = z \left(\log \frac{1}{z}\right)^3$ in the interval $z\in(0,1)$.
    The maximal value $\frac{27}{e^3}$ is located at $e^{-3}$.
    When $F(e^{-3})>F_0$,  it is clear that the line $F=F_0$  intersects the function's curve at two points, which correspond to two solutions, $z_1$ and $z_2$.}
\end{figure}

We demonstrate the existence of an island solution without explicitly solving for its position. Specifically, under the specified constraints and near-horizon conditions, we find that the entanglement entropy is approximately equal to the area entropy of the acoustic horizon
\bea
      S_{\rm gen} &=& \frac{2\pi a^2}{G_N} -  c\kappa\frac{2\pi b^2(r_+-r_h)}{r_+^3 \log\left(\frac{x}{y}\right)^2}  > \frac{2\pi a^2}{G_N}  -  c\kappa\frac{2\pi b^2(r_+-r_h)}{9r_+^3 } \nonumber \\
     &>&  \frac{2\pi r_+^2}{G_N}\left(1-c\kappa\frac{ G_Nb^2(r_+-r_h)}{9r_+^5 }\right) \nonumber \\
     &>& \frac{2\pi r_+^2}{G_N}\left(1-\frac{3}{2e^3}x^2\right)    \simeq \frac{2\pi r_+^2}{G_N}.
\eea
\subsection{The radiation region far from the horizon}
Unlike the case where the radiation region is close to the horizon, when the radiation region is far from the horizon, the position of the island can be directly solved. Using the formula Eq.~\eqref{eq:ee far formula}, one can derive the entanglement entropy of quantized phonons by taking the limit $\kappa_+\to 0$
\bea
S_{\rm phonon} &=& \lim_{\kappa_+\to 0} \frac{c}{3}\log\left\{\frac{2\sqrt{{\cal F}(a){\cal F}(b)}}{\kappa_+^2}\left[\cosh\left[\kappa_+\left({r_*(a)-r_*(b)}\right)\right]-\cosh\left[\kappa_+(t_a-t_b)\right]\right]\right\} \nonumber \\
    &=& \frac{c}{3}\log\left\{\sqrt{{\cal F}(a){\cal F}(b)}\left[\left({r_*(a)-r_*(b)}\right)^2-(t_a-t_b)^2\right]\right\}.
\eea
Then the generalized entropy is
\be
  S_{\rm gen} = \frac{2\pi a^2}{G_N}  +  \frac{c}{6}\log\left\{{\cal{F}}(a){\cal{F}}(b)\left[\left(r^*(a)-r^*(b)\right)^2 -\left(t_a-t_b\right)^2\right]^2\right\}.
\ee
Extremizing $S_{\rm gen}$ with respect to $t_a$ gives
\be
  \frac{\partial S_{\rm gen}}{\partial t_a} = \frac{2c\left(t_a-t_b\right)}{3\left[\left(t_a-t_b\right)^2+\left(r^*(a)-r^*(b)\right)^2\right]}=0,
\ee
which results $t_a=t_b$. Then extremizing the $S_{\rm gen}$ with respect to the location of island
\be
    \frac{\partial S_{\rm gen}}{\partial a} = \frac{4\pi a}{G_N} + \frac{c{\cal F}'(a)}{6{\cal F}(a)} +\frac{2c r^{*}(a)'}{3\left(r^*(a)-r^*(b)\right)}=0,
\ee
where ${\cal F}'(r)$ and $r^{*}(r)'$ are derivative of ${\cal F}(r)$ and $r^{*}(r)$, respectively. By taking the approximation $a\approx r_+$, one can finally obtain the location of the quantum extremal island as
\be
  a \simeq r_+ + \frac{c G_N}{6 \pi r_+}.
\ee
At late times, the generalized entanglement entropy with  island is given by
\be
  S_{\rm gen}  \simeq \frac{2\pi r_+^2}{G_N} +  \frac{c}{6}\log\left(\frac{36\pi r_+^{11}(b-r_+)^2(b-r_h)}{b^{3}c^2G_N^2(r_+-r_h)^{3}}\right)+ {\cal O}(G_N).
\ee
Note that it cannot be obtained from the continuous extremal limit of the non-extremal entanglement entropy in Eq.~(\ref{eq:far ee}) as we discussed before.

So far we have determined the contributions of islands to the entanglement entropy when the radiation region is either close to or far from the horizon. In both cases, the fine-grained entropy eventually reaches an upper bound. Furthermore, due to the symmetry of the Penrose diagram, the entanglement entropy of the radiation ultimately approaches twice the area entropy at the acoustic horizon.
\section{Page curve and Page time}\label{pagecurve}
\begin{figure}
    \includegraphics[width=0.7\textwidth]{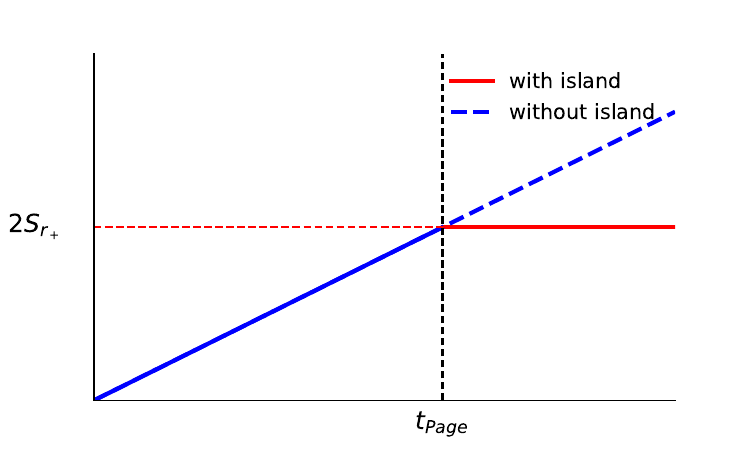}
    \caption{The Page curve of the entanglement entropy in the non-extremal case. The entanglement entropy is initially described by the result without islands, and then transitions to the case with islands after reaching the Page time. The entanglement entropy keep stable and could be approximated by the area entropy of the acoustic horizon.}
    \label{Figure:page}
\end{figure}

In this section, we will discuss the Page curve. To be precise, the entanglement entropy of radiation initially increases, reaches its maximum at the Page time, and then decreases gradually. In the island description, the entanglement entropy is initially described by the result without islands; as time progresses and after reaching the Page time, it transitions to the case with islands.
Thus the Page curve can be reproduced (see Figure \ref{Figure:page}).

In the non-extremal case, it can be expressed as:
\be
    S_{\rm gen} = {\rm Min}\left(\frac{c}{3}\kappa_+ t , S_{\rm gen}^{\text{with island}}\right),
\ee
where $S_{\rm gen}^{\text{with island}}$ is the entanglement entropy with islands, given by Eq.~\eqref{eq:close ee} and Eq.~\eqref{eq:far ee} depending on the location of the radiation region.
The Page time is therefore calculated as
\be
    t_{\rm Page}  =  \frac{ 6\pi r_+^2 }{G_N c \kappa_+} = \frac{6\pi r_h^3}{G_N c} \frac{ \xi \left(\xi + \sqrt{\left(\xi-4\right)\xi}\right)^2}{4-\xi + \sqrt{\left(\xi-4\right)\xi}}.
\ee
In the extremal limit where $\xi\to4$, we find that the Page time diverges, implying that the Page time cannot be well-defined. This is consistent with our previous discussion in Section \ref{extremal}. The Page time also diverges for the limit $\xi\to\infty$, where the acoustic horizon $r_+$ tends to infinity, meaning that the radiation cannot escape from the entire spacetime.

We have analyzed the Penrose diagram and calculated the entanglement entropy for the extremal case with an island, considering both close and far radiation regions. Many physical quantities diverge in this case due to the vanishing surface gravity $\kappa_+$, suggesting close connections between entanglement entropy and causal structure of spacetime. When the island is considered, the total entanglement entropy keeps stable again and is consistently equal to twice the area entropy of the acoustic horizon.

\section{Conclusion and discussion}\label{con}
In this paper, we investigated the information paradox of the acoustic black hole in a Schwarzschild black hole background using the island formula. By adopting different methods to estimate the entanglement entropy of phonons, distinct scenarios were considered: the radiation region is close to the acoustic horizon and is far from it. For the non-extremal case, we solved for the positions of islands and derived the entanglement entropy, which depends on the location of the radiation region. It was shown that the entanglement entropy can reach the steady value and the Page curve can be found only in the presence of island. When taking the extremal limit, the Page time will diverge, implying that the Page curve cannot be well defined in the extremal case. For the extremal case, our analysis showed that in the absence of islands, the entanglement entropy diverges due to the vanishing surface gravity, or zero Hawking-like temperature, which implied a close connection between entanglement entropy and spacetime causal structure. This finding is distinct from previous work on RN black holes, where the divergence was attributed to the inclusion of the black hole singularity in the island formula. 
Nevertheless, in the presence of islands, a finite entanglement entropy can be obtained, which to leading order is equal to twice the area entropy of the acoustic horizon.

Our results successfully reproduced the Page curve in this curved acoustic black hole system, confirming that unitarity is not violated in this analogue gravity model. This could provide new insights into both analogue gravity and real gravity. At present, we cannot directly observe the Hawking radiation, entanglement entropy, and the spacetime region known as the island of a real black hole. However, our results using the analogue gravity model suggested a resolution to this difficulty. For instance, in the model we adopted, the position of the acoustic horizon is affected by the flow velocity of the fluid surrounding the black hole and is always larger than that of the event horizon. This will result in the island region associated with phonon entanglement being exposed outside the event horizon, which means that we can observe the island region without having to access the interior of the black hole. Secondly, Our results indicated that the island formula may not only be applicable in the AdS/CFT correspondence, but also serve as a universal mechanism for restoring unitarity in any effective theory with a horizon. This is similar to the viewpoint in analogue gravity research, which asserts that Hawking-like radiation arises for any test field in any Lorentzian geometry containing an event horizon, without the need to satisfy the Einstein field equations \cite{Visser：1997yu}, and is also similar to the study of Hawking-Unurh effect in the noninertial reference frame~\cite{Huang:2007tw}. Thirdly, there have been also examples of discussing acoustic black holes in the context of holographic duality previously. For instance, \cite{Hossenfelder:2014gwa,Hossenfelder:2015pza} discussed how to use analog gravity systems to simulate AdS and asymptotically AdS spacetimes. \cite{Ge:2015uaa,Yu:2017bnu} constructed the duality between the phonon in the acoustic black hole and the quasinormal mode in the bulk perturbed AdS black brane through fluid/gravity duality.
The present work is a generalization of applying the holographic duality to the case of acoustic black holes, and we specifically investigated the properties of entanglement entropy in analogue gravity.

Several open questions remain in current research. For instance, it is unclear how to isolate the specific contributions of photons and phonons to the entanglement entropy in such a background, and how the resulting island structure compares to that of a standard Schwarzschild black hole. We leave these intriguing questions for future investigation. It is straightforward to generalize the calculation of entanglement entropy to other backgrounds, such as the RN black hole or AdS black holes. It is also worthwhile to use analogue gravity to realize the holographic principle.
\section*{Acknowledgement}
We would like to thank Xian-Hui Ge and Ruo-Han Wang for useful discussions. This work was supported by the National Natural Science Foundation of China (No.~12475069) and Guangdong Basic and Applied Basic Research Foundation (No.~2025A1515011321).






\begin{references}

\bibitem{Hawking:1976ra}
S.~W.~Hawking,
``Breakdown of Predictability in Gravitational Collapse,''
Phys.\ Rev.\ D \textbf{14}, 2460 (1976).

\bibitem{Hawking:1974rv}
S.~W.~Hawking,
``Black hole explosions,''
Nature \textbf{248}, 30 (1974).

\bibitem{Hawking:1975vcx}
S.~W.~Hawking,
``Particle Creation by Black Holes,''
Commun.\ Math.\ Phys. \textbf{43}, 199 (1975)
[erratum: Commun. Math. Phys. \textbf{46}, 206 (1976)].

\bibitem{Page:1993wv}
D.~N.~Page,
``Information in black hole radiation,''
Phys.\ Rev.\ Lett. \textbf{71}, 3743 (1993)
[arXiv:hep-th/9306083 [hep-th]].


\bibitem{Ryu:2006bv}
S.~Ryu and T.~Takayanagi,
``Holographic derivation of entanglement entropy from AdS/CFT,''
Phys.\ Rev.\ Lett. \textbf{96}, 181602 (2006)
[arXiv:hep-th/0603001 [hep-th]].

\bibitem{Hubeny:2007xt}
V.~E.~Hubeny, M.~Rangamani and T.~Takayanagi,
``A Covariant holographic entanglement entropy proposal,''
JHEP \textbf{07}, 062 (2007)
[arXiv:0705.0016 [hep-th]].

\bibitem{Faulkner:2013ana}
T.~Faulkner, A.~Lewkowycz and J.~Maldacena,
``Quantum corrections to holographic entanglement entropy,''
JHEP \textbf{11}, 074 (2013)
[arXiv:1307.2892 [hep-th]].

\bibitem{Penington:2019npb}
G.~Penington,
``Entanglement Wedge Reconstruction and the Information Paradox,''
JHEP \textbf{09}, 002 (2020)
[arXiv:1905.08255 [hep-th]].

\bibitem{Almheiri:2019psf}
A.~Almheiri, N.~Engelhardt, D.~Marolf and H.~Maxfield,
``The entropy of bulk quantum fields and the entanglement wedge of an evaporating black hole,''
JHEP \textbf{12}, 063 (2019)
[arXiv:1905.08762 [hep-th]].

\bibitem{Almheiri:2019hni}
A.~Almheiri, R.~Mahajan, J.~Maldacena and Y.~Zhao,
``The Page curve of Hawking radiation from semiclassical geometry,''
JHEP \textbf{03}, 149 (2020)
[arXiv:1908.10996 [hep-th]].

\bibitem{Almheiri:2020cfm}
A.~Almheiri, T.~Hartman, J.~Maldacena, E.~Shaghoulian and A.~Tajdini,
``The entropy of Hawking radiation,''
Rev.\ Mod.\ Phys. \textbf{93}, 035002 (2021)
[arXiv:2006.06872 [hep-th]].

\bibitem{Penington:2019kki}
G.~Penington, S.~H.~Shenker, D.~Stanford and Z.~Yang,
``Replica wormholes and the black hole interior,''
JHEP \textbf{03}, 205 (2022)
[arXiv:1911.11977 [hep-th]].

\bibitem{Almheiri:2019qdq}
A.~Almheiri, T.~Hartman, J.~Maldacena, E.~Shaghoulian and A.~Tajdini,
``Replica Wormholes and the Entropy of Hawking Radiation,''
JHEP \textbf{05}, 013 (2020)
[arXiv:1911.12333 [hep-th]].

\bibitem{Yu:2025tid}
M.~H.~Yu, S.~Y.~Lin and X.~H.~Ge,
``Replica Wormholes, Modular Entropy, and Capacity of Entanglement in JT Gravity,''
[arXiv:2501.11474 [hep-th]].

\bibitem{Bombelli:1986rw}
L.~Bombelli, R.~K.~Koul, J.~Lee and R.~D.~Sorkin,
``A Quantum Source of Entropy for Black Holes,''
Phys.\ Rev.\ D \textbf{34}, 373 (1986).

\bibitem{Srednicki:1993im}
M.~Srednicki,
``Entropy and area,''
Phys.\ Rev.\ Lett. \textbf{71}, 666 (1993)
[arXiv:hep-th/9303048 [hep-th]].

\bibitem{Susskind:1994sm}
L.~Susskind and J.~Uglum,
``Black hole entropy in canonical quantum gravity and superstring theory,''
Phys.\ Rev.\ D \textbf{50}, 2700 (1994)
[arXiv:hep-th/9401070 [hep-th]].

\bibitem{Hollowood:2020cou}
T.~J.~Hollowood and S.~P.~Kumar,
``Islands and Page Curves for Evaporating Black Holes in JT Gravity,''
JHEP \textbf{08}, 094 (2020)
[arXiv:2004.14944 [hep-th]].

\bibitem{Goto:2020wnk}
K.~Goto, T.~Hartman and A.~Tajdini,
``Replica wormholes for an evaporating 2D black hole,''
JHEP \textbf{04}, 289 (2021)
[arXiv:2011.09043 [hep-th]].

\bibitem{Anegawa:2020ezn}
T.~Anegawa and N.~Iizuka,
``Notes on islands in asymptotically flat 2d dilaton black holes,''
JHEP \textbf{07}, 036 (2020)
[arXiv:2004.01601 [hep-th]].

\bibitem{Gautason:2020tmk}
F.~F.~Gautason, L.~Schneiderbauer, W.~Sybesma and L.~Thorlacius,
``Page Curve for an Evaporating Black Hole,''
JHEP \textbf{05}, 091 (2020)
[arXiv:2004.00598 [hep-th]].

\bibitem{Hartman:2020swn}
T.~Hartman, E.~Shaghoulian and A.~Strominger,
``Islands in Asymptotically Flat 2D Gravity,''
JHEP \textbf{07}, 022 (2020)
[arXiv:2004.13857 [hep-th]].

\bibitem{Wang:2021mqq}
X.~Wang, R.~Li and J.~Wang,
``Page curves for a family of exactly solvable evaporating black holes,''
Phys.\ Rev.\ D \textbf{103}, 126026 (2021)
[arXiv:2104.00224 [hep-th]].

\bibitem{Wang1:2021woy}
X.~Wang, R.~Li and J.~Wang,
``Islands and Page curves of Reissner-Nordstr\"om black holes,''
JHEP \textbf{04}, 103 (2021)
[arXiv:2101.06867 [hep-th]].

\bibitem{Li:2021lfo}
R.~Li, X.~Wang and J.~Wang,
``Island may not save the information paradox of Liouville black holes,''
Phys.\ Rev.\ D \textbf{104}, 106015 (2021)
[arXiv:2105.03271 [hep-th]].

\bibitem{Almheiri:2019psy}
A.~Almheiri, R.~Mahajan and J.~E.~Santos,
``Entanglement islands in higher dimensions,''
SciPost Phys. \textbf{9}, 001 (2020)
[arXiv:1911.09666 [hep-th]].

\bibitem{He:2021mst}
S.~He, Y.~Sun, L.~Zhao and Y.~X.~Zhang,
``The universality of islands outside the horizon,''
JHEP \textbf{05}, 047 (2022)
[arXiv:2110.07598 [hep-th]].

\bibitem{Alishahiha:2020qza}
M.~Alishahiha, A.~Faraji Astaneh and A.~Naseh,
``Island in the presence of higher derivative terms,''
JHEP \textbf{02}, 035 (2021)
[arXiv:2005.08715 [hep-th]].

\bibitem{Hashimoto:2020cas}
K.~Hashimoto, N.~Iizuka and Y.~Matsuo,
``Islands in Schwarzschild black holes,''
JHEP \textbf{06}, 085 (2020)
[arXiv:2004.05863 [hep-th]].

\bibitem{Krishnan:2020oun}
C.~Krishnan, V.~Patil and J.~Pereira,
``Page Curve and the Information Paradox in Flat Space,''
[arXiv:2005.02993 [hep-th]].

\bibitem{Hu:2022zgy}
P.~J.~Hu, D.~Li and R.~X.~Miao,
``Island on codimension-two branes in AdS/dCFT,''
JHEP \textbf{11}, 008 (2022)
[arXiv:2208.11982 [hep-th]].

\bibitem{Gan:2022jay}
W.~C.~Gan, D.~H.~Du and F.~W.~Shu,
``Island and Page curve for one-sided asymptotically flat black hole,''
JHEP \textbf{07}, 020 (2022)
[arXiv:2203.06310 [hep-th]].

\bibitem{Du:2022vvg}
D.~H.~Du, W.~C.~Gan, F.~W.~Shu and J.~R.~Sun,
``Unitary constraints on semiclassical Schwarzschild black holes in the presence of island,''
Phys.\ Rev.\ D \textbf{107}, 026005 (2023)
[arXiv:2206.10339 [hep-th]].

\bibitem{Ling:2020laa}
Y.~Ling, Y.~Liu and Z.~Y.~Xian,
``Island in Charged Black Holes,''
JHEP \textbf{03}, 251 (2021)
[arXiv:2010.00037 [hep-th]].

\bibitem{Wang:2021woy}
X.~Wang, R.~Li and J.~Wang,
``Islands and Page curves of Reissner-Nordstr\"om black holes,''
JHEP \textbf{04}, 103 (2021)
[arXiv:2101.06867 [hep-th]].

\bibitem{Miao:2023unv}
R.~X.~Miao,
``Entanglement island and Page curve in wedge holography,''
JHEP \textbf{03}, 214 (2023)
[arXiv:2301.06285 [hep-th]].

\bibitem{Guo:2023fly}
Y.~Guo and R.~X.~Miao,
``Page curves on codim-m and charged branes,''
Eur.\ Phys.\ J.\ C \textbf{83}, 847 (2023).

\bibitem{Guo:2023gfa}
C.~Z.~Guo, W.~C.~Gan and F.~W.~Shu,
``Page curves and entanglement islands for the step-function Vaidya model of evaporating black holes,''
JHEP \textbf{05}, 042 (2023)
[arXiv:2302.02379 [hep-th]].

\bibitem{Kim:2021gzd}
W.~Kim and M.~Nam,
``Entanglement entropy of asymptotically flat non-extremal and extremal black holes with an island,''
Eur.\ Phys.\ J.\ C \textbf{81}, 869 (2021)
[arXiv:2103.16163 [hep-th]].

\bibitem{Tong:2023nvi}
C.~W.~Tong, D.~H.~Du and J.~R.~Sun,
``Island of Reissner-Nordstr\"om anti-de Sitter black holes in the large D limit,''
Phys.\ Rev.\ D \textbf{109}, 104053 (2024)
[arXiv:2306.06682 [hep-th]].

\bibitem{Du:2025kcx}
Y.~Du, J.~R.~Sun and X.~Zhang,
``Information paradox and island of covariant black holes in LQG,''
[arXiv:2510.11921 [gr-qc]].

\bibitem{Ahn:2021chg}
B.~Ahn, S.~E.~Bak, H.~S.~Jeong, K.~Y.~Kim and Y.~W.~Sun,
``Islands in charged linear dilaton black holes,''
Phys.\ Rev.\ D \textbf{105}, 046012 (2022)
[arXiv:2107.07444 [hep-th]].

\bibitem{HosseiniMansoori:2022hok}
S.~A.~Hosseini Mansoori, O.~Luongo, S.~Mancini, M.~Mirjalali, M.~Rafiee and A.~Tavanfar,
``Planar black holes in holographic axion gravity: Islands, Page times, and scrambling times,''
Phys.\ Rev.\ D \textbf{106}, 126018 (2022)
[arXiv:2209.00253 [hep-th]].

\bibitem{RoyChowdhury:2022awr}
A.~Roy Chowdhury, A.~Saha and S.~Gangopadhyay,
``Role of mutual information in the Page curve,''
Phys. Rev. D \textbf{106} (2022) no.8, 086019
[arXiv:2207.13029 [hep-th]].

\bibitem{Yu:2024fks}
M.~H.~Yu and X.~H.~Ge,
``Geometric constraints via Page curves: insights from island rule and quantum focusing conjecture*,''
Chin. Phys. C \textbf{49} (2025) no.4, 045107
[arXiv:2405.03220 [hep-th]].

\bibitem{Espindola:2022fqb}
R.~Esp{\'\i}ndola, B.~Najian and D.~Nikolakopoulou,
``Islands in FRW Cosmologies,''
[arXiv:2203.04433 [hep-th]].

\bibitem{Barcelo:2005fc}
C.~Barcelo, S.~Liberati and M.~Visser,
``Analogue gravity,''
Living Rev. Rel. \textbf{8}, 12 (2005)
[arXiv:gr-qc/0505065 [gr-qc]].

\bibitem{Unruh:1980cg}
W.~G.~Unruh,
``Experimental black hole evaporation,''
Phys.\ Rev.\ Lett. \textbf{46}, 1351 (1981).

\bibitem{Lahav:2009wx}
O.~Lahav, A.~Itah, A.~Blumkin, C.~Gordon and J.~Steinhauer,
``Realization of a sonic black hole analogue in a Bose-Einstein condensate,''
Phys.\ Rev.\ Lett. \textbf{105}, 240401 (2010)
[arXiv:0906.1337 [cond-mat.quant-gas]].

\bibitem{Drori:2018ivu}
J.~Drori, Y.~Rosenberg, D.~Bermudez, Y.~Silberberg and U.~Leonhardt,
``Observation of Stimulated Hawking Radiation in an Optical Analogue,''
Phys.\ Rev.\ Lett. \textbf{122}, 010404 (2019)
[arXiv:1808.09244 [gr-qc]].

\bibitem{Steinhauer:2015saa}
J.~Steinhauer,
``Observation of quantum Hawking radiation and its entanglement in an analogue black hole,''
Nature Phys. \textbf{12}, 959 (2016)
[arXiv:1510.00621 [gr-qc]].

\bibitem{Fedichev:2003id}
P.~O.~Fedichev and U.~R.~Fischer,
``Gibbons-Hawking effect in the sonic de Sitter space-time of an expanding Bose-Einstein-condensed gas,''
Phys. Rev. Lett. \textbf{91} (2003), 240407
[arXiv:cond-mat/0304342 [cond-mat]].


\bibitem{Fischer:2004bf}
U.~R.~Fischer and R.~Schutzhold,
``Quantum simulation of cosmic inflation in two-component Bose-Einstein condensates,''
Phys. Rev. A \textbf{70} (2004), 063615
[arXiv:cond-mat/0406470 [cond-mat.other]].

\bibitem{Cha:2016esj}
S.~Y.~Ch{\"a} and U.~R.~Fischer,
``Probing the scale invariance of the inflationary power spectrum in expanding quasi-two-dimensional dipolar condensates,''
Phys. Rev. Lett. \textbf{118} (2017) no.13, 130404
[arXiv:1609.06155 [cond-mat.quant-gas]].

\bibitem{Ge:2015uaa}
X.~H.~Ge, J.~R.~Sun, Y.~Tian, X.~N.~Wu and Y.~L.~Zhang,
``Holographic Interpretation of Acoustic Black Holes,''
Phys.\ Rev.\ D \textbf{92}, 084052 (2015)
[arXiv:1508.01735 [hep-th]].

\bibitem{Yu:2017bnu}
C.~Yu and J.~R.~Sun,
``Note on acoustic black holes from black D3-brane,''
Int.\ J.\ Mod.\ Phys.\ D \textbf{28}, 1950095 (2019)
[arXiv:1712.04137 [hep-th]].




\bibitem{Ge:2019our}
X.~H.~Ge, M.~Nakahara, S.~J.~Sin, Y.~Tian and S.~F.~Wu,
``Acoustic black holes in curved spacetime and the emergence of analogue Minkowski spacetime,''
Phys.\ Rev.\ D \textbf{99}, 104047 (2019)
[arXiv:1902.11126 [hep-th]].


\bibitem{Das:2004zm}
T.~K.~Das,
``Analogue Hawking radiation from astrophysical black hole accretion,''
Class. Quant. Grav. \textbf{21}, 5253-5260 (2004)
[arXiv:gr-qc/0408081 [gr-qc]].


\bibitem{Wang:2019zqw}
Q.~B.~Wang and X.~H.~Ge,
``Geometry outside of acoustic black holes in (2+1)-dimensional spacetime,''
Phys.\ Rev.\ D \textbf{102}, 104009 (2020)
[arXiv:1912.05285 [hep-th]].

\bibitem{Guo:2020blq}
H.~Guo, H.~Liu, X.~M.~Kuang and B.~Wang,
``Acoustic black hole in Schwarzschild spacetime: quasi-normal modes, analogous Hawking radiation and shadows,''
Phys.\ Rev.\ D \textbf{102}, 124019 (2020)
[arXiv:2007.04197 [gr-qc]].

\bibitem{Ling:2021vgk}
R.~Ling, H.~Guo, H.~Liu, X.~M.~Kuang and B.~Wang,
``Shadow and near-horizon characteristics of the acoustic charged black hole in curved spacetime,''
Phys.\ Rev.\ D \textbf{104}, 104003 (2021)
[arXiv:2107.05171 [gr-qc]].

\bibitem{Qiao:2021trw}
C.~K.~Qiao and M.~Zhou,
``The gravitational bending of acoustic Schwarzschild black hole,''
Eur.\ Phys.\ J.\ C \textbf{83}, 271 (2023)
[arXiv:2109.05828 [gr-qc]].

\bibitem{Wang:2022gbl}
Q.~B.~Wang, M.~H.~Yu and X.~H.~Ge,
``Scrambling time for analogue black holes embedded in AdS space,''
Eur.\ Phys.\ J.\ C \textbf{82}, 468 (2022)
[arXiv:2203.07914 [hep-th]].

\bibitem{Parvizi:2023foz}
S.~Parvizi and M.~Shahbazi,
``Analogue gravity and the island prescription,''
Eur.\ Phys.\ J.\ C \textbf{83}, 705 (2023)
[arXiv:2302.08742 [hep-th]].

\bibitem{Casini:2005zv}
H.~Casini and M.~Huerta,
``Entanglement and alpha entropies for a massive scalar field in two dimensions,''
J.\ Stat.\ Mech. \textbf{0512}, P12012 (2005)
[arXiv:cond-mat/0511014 [cond-mat]].

\bibitem{Casini:2009sr}
H.~Casini and M.~Huerta,
``Entanglement entropy in free quantum field theory,''
J.\ Phys.\ A \textbf{42}, 504007 (2009)
[arXiv:0905.2562 [hep-th]].

\bibitem{Visser：1997yu}
M.~Visser,
``Hawking radiation without black hole entropy,''
Phys.\ Rev.\ Lett. \textbf{80}, 3436--3439 (1998)
[arXiv:9712016 [gr-qc]].


\bibitem{Huang:2007tw}
C.~G.~Huang and J.~R.~Sun,
``Thermodynamic properties of spherically-symmetric, uniformly-accelerated reference frames,''
Commun. Theor. Phys. \textbf{49}, 928-932 (2008)
[arXiv:gr-qc/0701078 [gr-qc]].

\bibitem{Hossenfelder:2014gwa}
S.~Hossenfelder,
``Analog Systems for Gravity Duals,''
Phys.\ Rev.\ D \textbf{91}, 124064 (2015)
[arXiv:1412.4220 [gr-qc]].

\bibitem{Hossenfelder:2015pza}
S.~Hossenfelder,
``A relativistic acoustic metric for planar black holes,''
Phys.\ Lett.\ B \textbf{752}, 13 (2016)
[arXiv:1508.00732 [gr-qc]].




\end{references}
\end{document}